\documentclass{aa}
\usepackage[utf8]{inputenc}
\usepackage[T1]{fontenc}
\usepackage{natbib}
\usepackage{amsmath}
\usepackage{amssymb}
\usepackage{geometry}
\usepackage{graphicx}
\usepackage{bm}
\usepackage{indentfirst}
\usepackage{hyperref}
\usepackage{caption}
\usepackage{subfig}
\usepackage{xcolor}
\usepackage{txfonts}
\usepackage{ulem}
\bibpunct{(}{)}{;}{a}{}{,}
\newcommand*\diff{\mathop{}\!\mathrm{d}}
\newcommand{\vv}{\mathrm{v}}
\usepackage{cleveref}

\begin{document}

\title{Velocity-intensity asymmetry reversal of solar radial $p$-modes}

\author{J. Philidet\inst{\ref{inst1}}}

\institute{LESIA, Observatoire de Paris, PSL Research University, CNRS, Universit\'e Pierre et Marie Curie, Universit\'e Paris Diderot, 92195 Meudon, France \label{inst1}}

\author{J. Philidet\inst{\ref{inst1}} \and K. Belkacem\inst{\ref{inst1}} \and H.-G. Ludwig\inst{\ref{inst2},\ref{inst3}} \and R. Samadi\inst{\ref{inst1}} \and C. Barban\inst{\ref{inst1}}}

\institute{LESIA, Observatoire de Paris, PSL Research University, CNRS, Universit\'e Pierre et Marie Curie, Universit\'e Paris Diderot, 92195 Meudon, France \label{inst1} \and Zentrum für Astronomie der Universität Heidelberg, Landessternwarte, Königstuhl 12, 69117 Heidelberg, Germany\label{inst2} \and GEPI, Observatoire de Paris, PSL Research University, CNRS, Universit\'e Pierre et Marie Curie, Universit\'e Paris Diderot, 92195 Meudon, France\label{inst3}}

\abstract{
The development of space-borne missions has significantly improved the quality of the measured spectrum of solar-like oscillators. Their $p$-mode line profiles can now be resolved, and their asymmetry inferred for a variety of stars other than the Sun. However, it has been known for a long time that the asymmetries of solar $p$-modes are reversed between the velocity and the intensity spectra. Understanding the origin of this reversal is necessary in order to use asymmetries as a tool for seismic diagnosis. Indeed, for stars other than the Sun, only the intensity power spectrum is sufficiently resolved to allow for an estimation of mode asymmetries. In \citet{philidet20}, we developed an approach designed to model and predict these asymmetries in the velocity power spectrum of the Sun and to successfully compare them to their observationally derived counterpart. In this paper, we expand our model and predict the asymmetries featured in the intensity power spectrum. We find that the shape of the mode line profiles in intensity is largely dependent on how the oscillation-induced variations of the radiative flux are treated, and that modelling it realistically is crucial for understanding asymmetry reversal. Perturbing a solar-calibrated grey atmosphere model, and adopting the quasi-adiabatic framework as a first step, we reproduce the asymmetries observed in solar intensity spectra for low-frequency modes. We conclude that, unlike what was previously thought, it is not necessary to invoke an additional mechanism (e.g non-adiabatic effects, coherent non-resonant background signal) to explain asymmetry reversal. Such an additional mechanism is necessary, however, to explain asymmetry reversal for higher-order modes.
}


\maketitle

\section{Introduction}

The power spectral density of solar-like oscillations is expected to feature Lorentzian-shaped peaks centered on their eigenfrequencies. However, observations show that their line profiles are slightly asymmetric \citep{duvall93}. This asymmetry is primarily due to the fact that stochastic excitation occurs in a localised region just beneath the surface of the star. Several studies have consequently used observed line profile asymmetries to infer properties of the turbulent source of excitation of solar $p$-modes, in particular its radial position and its multipolar nature \citep[see for instance][]{roxburghV97,nigam98}. These prior studies used parameterised models, and aimed to find best-fit values for their free parameters by applying fitting procedures to individual peaks in the observed spectrum. \citet{philidet20} followed a different approach, which consisted in modelling mode asymmetry without fitting any free parameters to the available observational data. Instead, they developed an analytical model of stochastic excitation, coupled with a 3D hydrodynamical simulation of the stellar atmosphere. This allowed the authors to reproduce the asymmetries of solar $p$-modes as measured in the observed velocity power spectrum, and to subsequently demonstrate the role of the spatial extent of the mode driving region, together with the differential properties of turbulent convection (namely the variation of the injection length-scale below and above the photosphere).

The recent measurement of $p$-mode asymmetries in solar-like oscillators by \citet{benomar18}, using \textit{Kepler} observations, opened the way for constraining the properties of stochastic excitation, and therefore of turbulent convection, throughout the HR diagram. Unlike in the solar case, however, only observations made by \textit{Kepler} in intensity have the resolution necessary for mode asymmetry to be inferred. But it is well known that asymmetries are very different between velocity and intensity observations \citep{duvall93,toutain97}. More precisely, line profiles in the velocity power spectrum have more power on their low-frequency side, whereas line profiles in the intensity power spectrum have more power in their high-frequency side. This is known as the asymmetry reversal puzzle.

Several hypotheses have been proposed to explain this reversal, but no consensus emerged. \citet{duvall93} suggested that modal entropy fluctuations could affect intensity asymmetries while leaving asymmetries in velocity unaffected. Following this argument, \citet{rastB98} quantified these non-adiabatic effects, and found that radiative cooling has but a negligible impact on line profile asymmetry, so that it can hardly explain the asymmetry reversal. \citet{gabriel98} then remarked that in order for such heat transfers to impact line asymmetry, it is necessary that they affect the wave equation beyond the second order, which was not the case in the model of \citet{rastB98}.

Shortly afterwards, another candidate was found to explain the asymmetry reversal puzzle. \citet{roxburghV97} remarked that adding a coherent, non-resonant background (i.e a signal with a very broad spectrum, but whose Fourier component close to an oscillating mode would be coherent with it, and would therefore be able to interfere with it) to a resonant signal could affect its line profile, and thus its asymmetry. They argued that the overshooting of turbulent eddies into the lower layers of the atmosphere should act as a coherent non-resonant background in the velocity signal, while keeping the intensity signal unaffected, thus explaining the asymmetry reversal. On the other hand, \citet{nigam98} argued that both signals contain a coherent background component, and that the asymmetry reversal stems from the fact that it is much stronger in the intensity signal than in the velocity signal. Subsequently, several studies have undertaken the task of estimating the level of correlated background in the intensity signal using the observed asymmetry reversal \citep[e.g][]{kumarB99,chaplinA99,severino01,barban04}.

Yet another candidate to explain asymmetry reversal was recognised by \citet{georgo03}. They noticed that $p$-modes featured in the velocity and intensity power spectra of 3D hydrodynamic simulations of the stellar uppermost layers have opposite asymmetries when they are computed at fixed optical depth, but identical asymmetries when they are computed at fixed geometrical depth. They therefore proposed the following picture: the velocity and intensity line profiles are intrinsically proportional to one another. But the height of unity optical depth, at which the intensity fluctuations are observed, oscillates with the mode. Since there is a strong temperature gradient there, the \textit{observed} background temperature also oscillates, which adds a component to the resonant intensity signal. They showed that this added component tends to reduce the amplitude of the intensity fluctuations. But, because the $\kappa-T$ relation is highly non-linear at the photosphere, this reduction is not uniform over the entire line profile: as such, it changes its asymmetry. Since there is no equilibrium velocity gradient, there is no corresponding additional component in the velocity signal, and the velocity asymmetry remains unaffected.

At the core of the aforementioned models lies the assumption that the gas temperature at the photosphere and the effective temperature are equal, and therefore have equal relative fluctuations. From this assumption stems the fact that velocity and intensity line profiles should have the same asymmetry, so that the key to understand the asymmetry reversal puzzle must be sought elsewhere. While this assumption is not so problematic when it is used as a photospheric boundary condition in the framework of fully non-adiabatic calculations \citep{dupret02}, it becomes more questionable when used to justify that luminosity and temperature eigenfunctions are simply proportional to one another. In this context, our objective is to question this assumption, and to show that the radiative flux reacts more complexly to temperature variations, which has a crucial impact on line profile asymmetry in the intensity spectra. To that end, we extend the model of \citet{philidet20} for intensity observations. We show that, unlike what was previously thought, source localisation impacts velocity and intensity observations differently, so that an additional physical mechanism is not necessary to account for asymmetry reversal, except maybe at high frequency. This paper is structured as follows: in Sect. \ref{sec:model}, we describe the steps necessary for the adaptation of our model to intensity observations; we then present its predictions pertaining to asymmetry reversal in Sect. \ref{sec:obs}, and discuss these results in Sect. \ref{sec:discussion}.

\section{Modelling the intensity power spectrum\label{sec:model}}

In order to adapt our $p$-mode stochastic excitation model to intensity observations, it is first necessary to model the intensity Green's function associated to radial $p$-modes. Indeed, in \citet{philidet20}, the wave equation was written in terms of a wave variable $\Psi_\omega$, related to the velocity fluctuations through
\begin{equation}
\widehat{\vv_\text{osc}}(r) = \dfrac{j\omega}{rc\sqrt{\rho}}\Psi_\omega(r)~,
\label{eq:v-psi}
\end{equation}
where $\rho$ is the equilibrium density, $c$ is the sound speed, $\omega$ the angular frequency, $j$ the imaginary unit, and the notation $\widehat{.}$ refers to Fourier transform in time. The radial inhomogeneous wave equation then reads
\begin{equation}
\dfrac{\diff^2\Psi_\omega}{\diff r^2} + \left(\dfrac{\omega^2 + j\omega\Gamma_\omega}{c^2} - V(r)\right)\Psi_\omega = S(r)~,
\label{eq:wave_eq}
\end{equation}
where $V(r)$ is the acoustic potential, $\Gamma_\omega$ is the frequency-dependent damping rate of the oscillations, and $S(r)$ is a source term proportional to the divergence of the fluctuating Reynolds stresses. The damping rates $\Gamma_\omega$ are inferred from observations; we use the same values as those presented in Table 1 of \citet{philidet20}.

For a given value of $\omega$, convolving the Green's function of Eq. \eqref{eq:wave_eq} with the source term $S(r)$ allows us to predict the value of the power spectrum in terms of $\Psi_\omega$. Then the velocity power spectrum is given by Eq. \eqref{eq:v-psi}. Following the same approach, the goal of this section is to relate the intensity fluctuations to the wave variable $\Psi_\omega$, so that from the $\Psi_\omega$-power spectrum we may have access to the intensity power spectrum.

\subsection{Intensity fluctuations in a grey atmosphere\label{sec:grey_atm}}

Because the $p$-mode intensity power spectrum is observed close to the photosphere, we only need to model the oscillation-induced intensity variations in this region. In order to model these variations, we will treat the atmosphere as a grey atmosphere. This is justified, in part, by the fact that the observed intensity power spectrum corresponds to the continuum intensity, on which absorption spectral lines have little impact.

A grey atmosphere is, by definition, in radiative equilibrium. When studying $p$-mode-driven intensity fluctuations, working under the radiative equilibrium assumption is justified by the large gap existing between the modal period and the local thermal timescale. The period of the modes is of the order of several minutes, while the thermal timescale is about ten seconds at the photosphere. As such, the radiative flux reacts almost instantly to the oscillations induced by the $p$-modes, and the atmosphere remains at radiative equilibrium at all time. One of the consequences of radiative equilibrium is that the radiative flux is uniform, and its relative fluctuations are proportional to the relative fluctuations of the effective temperature $T_\text{eff}$. The equality between the relative luminosity fluctuations at the photosphere and at the observation height of the modes is further supported by the fact that the modes are observed very close to the photosphere. The goal of the present section is therefore to relate the variations of $T_\text{eff}$ to the gas fluctuations brought about by the oscillating modes, and in particular to the gas temperature fluctuations. In doing so, we follow the same treatment of $p$-mode-induced atmospheric perturbation as \citet{dupret02}.

In a grey atmosphere, the temperature of the gas is given by a unique function of the optical depth $\tau$ and effective temperature $T_\text{eff}$, expressed by means of the Hopf function $q(\tau)$ \citep{mihalasbook}
\begin{equation}
T(\tau,T_\text{eff}) = \left(\dfrac{3}{4}(\tau + q(\tau))\right)^{1/4}T_\text{eff}~.
\label{eq:grey}
\end{equation}
An arbitrary relation $T = T(\tau, T_\text{eff})$ can be perturbed thusly
\begin{equation}
\dfrac{\delta T}{T} = \left(\dfrac{\partial \ln T}{\partial \ln T_\text{eff}}\right)_{\tau}\dfrac{\delta T_\text{eff}}{T_\text{eff}} + \left(\dfrac{\partial \ln T}{\partial \ln \tau}\right)_{T_\text{eff}}\dfrac{\delta \tau}{\tau}~,
\label{eq:deltaTsurT}
\end{equation}
where the fluctuations of the optical depth can be expressed as
\begin{equation}
\dfrac{\diff \delta \tau}{\diff \tau} = \dfrac{\delta \kappa}{\kappa} + \dfrac{\delta \rho}{\rho} + \dfrac{\partial \xi_r}{\partial r}~,
\label{eq:deltatausurtau}
\end{equation}
where $\kappa$ is the Rosseland mean opacity, and $\xi_r$ the radial displacement of the gas.

Taking the partial derivative of Eq. \eqref{eq:deltaTsurT} with respect to $\tau$, and eliminating $\diff\delta\tau / \diff\tau$ alternatively through Eqs. \eqref{eq:deltaTsurT} and \eqref{eq:deltatausurtau}, one finds
\begin{multline}
\dfrac{\partial (\delta T / T)}{\partial \ln\tau} = \dfrac{\partial \ln T}{\partial \ln\tau}\left(\dfrac{\delta \kappa}{\kappa} + \dfrac{\delta \rho}{\rho} + \dfrac{\partial \xi_r}{\partial r}\right) \\
- \left(1 - \dfrac{\partial^2 \ln T / \partial \ln\tau^2}{\partial \ln T / \partial \ln\tau}\right)\left(\dfrac{\delta T}{T} - \dfrac{\partial \ln T}{\partial \ln T_\text{eff}}\dfrac{\delta T_\text{eff}}{T_\text{eff}}\right) \\
+ \dfrac{\partial^2 \ln T}{\partial\ln\tau\partial\ln T_\text{eff}}\dfrac{\delta T_\text{eff}}{T_\text{eff}}~.
\label{eq:partialT}
\end{multline}

Isolating the perturbation of the effective temperature, this can be rearranged
\begin{multline}
\dfrac{\delta T_\text{eff}}{T_\text{eff}} = \left(\dfrac{\partial^2 \ln T}{\partial\ln\tau\partial\ln T_\text{eff}} + \dfrac{\partial \ln T}{\partial \ln T_\text{eff}}\left(1 - \dfrac{\partial^2 \ln T / \partial \ln\tau^2}{\partial \ln T / \partial \ln\tau}\right)\right)^{-1} \\
\times \left[\dfrac{\partial (\delta T / T)}{\partial \ln\tau} - \dfrac{\partial \ln T}{\partial \ln\tau}\left(\dfrac{\delta \kappa}{\kappa} + \dfrac{\delta \rho}{\rho} + \dfrac{\partial \xi_r}{\partial r}\right)\right. \\
\left. + \dfrac{\delta T}{T}\left(1 - \dfrac{\partial^2 \ln T / \partial \ln\tau^2}{\partial \ln T / \partial \ln\tau}\right)\right]~.
\label{eq:7}
\end{multline}

When the temperature law as given by Eq. \eqref{eq:grey} is used, Eq. \eqref{eq:7} reduces to
\begin{multline}
\dfrac{\delta T_\text{eff}}{T_\text{eff}} = \left(\dfrac{\tau r'(\tau)}{r(\tau)} - \dfrac{\tau r''(\tau)}{r'(\tau)} \right)^{-1}\left[\dfrac{\partial(\delta T / T)}{\partial \ln\tau} \right.\\
\left. - \dfrac{\tau r'(\tau)}{4r(\tau)}\left(\dfrac{\delta\kappa}{\kappa} + \dfrac{\delta\rho}{\rho} + \dfrac{\partial\xi_r}{\partial r}\right) + \dfrac{\delta T}{T}\left(\dfrac{\tau r'(\tau)}{r(\tau)} - \dfrac{\tau r''(\tau)}{r'(\tau)}\right)\right]~,
\label{eq:8}
\end{multline}
where we have introduced
\begin{equation}
r(\tau) \equiv \tau + q(\tau)~,
\end{equation}
and $'$ and $''$ respectively refer to the first and second derivatives with respect to $\tau$. The first term in the brackets can be rearranged as such
\begin{equation}
\dfrac{\partial (\delta T/T)}{\partial\ln\tau} = \dfrac{\tau r'(\tau)}{4r(\tau)}\left(\dfrac{\partial\delta T / \partial r}{\diff T / \diff r} - \dfrac{\delta T}{T}\right)~.
\label{eq:10}
\end{equation}
Finally, using Eqs. \eqref{eq:8} and \eqref{eq:10}, the fluctuations of the effective temperature can be written
\begin{multline}
\dfrac{\delta T_\text{eff}}{T_\text{eff}} = \dfrac{1}{4}\left(1 - \dfrac{r(\tau)r''(\tau)}{r'(\tau)^2}\right)^{-1} \left[\left(3 - \dfrac{r(\tau)r''(\tau)}{r'(\tau)^2}\right)\dfrac{\delta T}{T} \right. \\
\left. + \dfrac{\partial \delta T / \partial r}{\diff T / \diff r} - \dfrac{\delta\kappa}{\kappa} - \dfrac{\delta\rho}{\rho} - \dfrac{\partial\xi_r}{\partial r}\right]~.
\label{eq:deltaTeff}
\end{multline}

The fluctuations of the effective temperature, by essence, do not depend on $\tau$, and therefore, neither does the right-hand side of Eq. \eqref{eq:deltaTeff}, even though its individual terms do. The atmosphere being in radiative equilibrium, and the height of observation being fairly close to the photosphere, the radiative flux is uniform and equal to its value at the photosphere. This also holds true for the fluctuations of the radiative luminosity, which, by definition of the effective temperature, are therefore given by
\begin{equation}
\dfrac{\delta L_R}{L_R} = 4\dfrac{\delta T_\text{eff}}{T_\text{eff}} + 2\dfrac{\xi_r}{r}~,
\label{eq:rad_eq}
\end{equation}
which, thanks to Eq. \eqref{eq:deltaTeff}, becomes:
\begin{multline}
\dfrac{\delta L_R}{L_R} = 2\dfrac{\xi_r}{r} + \dfrac{1}{1-x(\tau)} \left[\left(3 - x(\tau)\right)\dfrac{\delta T}{T} \right. \\
\left. + \dfrac{\partial \delta T / \partial r}{\diff T / \diff r} - \dfrac{\delta\kappa}{\kappa} - \dfrac{\delta\rho}{\rho} - \dfrac{\partial\xi_r}{\partial r}\right]~.
\end{multline}
where we have introduced
\begin{equation}
x(\tau) \equiv \dfrac{r(\tau)r''(\tau)}{r'(\tau)^2} = \dfrac{(\tau + q(\tau))q''(\tau)}{(1+q(\tau))^2}
\label{eq:xtau}
\end{equation}

Finally, we neglect the Lagrangian perturbations of the convective luminosity $L_C$ compared to those of the radiative luminosity $L_R$. Indeed, including the convective contribution in this model would require a non-local, time-dependent treatment of turbulent convection; furthermore, the oscillations would have to be treated in a non-adiabatic framework. Until subsequent refinements are made to this model, and as a first approximation, we therefore consider $\delta L = \delta L_R$, in which case we have
\begin{multline}
\dfrac{\delta L}{L} = \dfrac{L_R}{L}\left(2\dfrac{\xi_r}{r} + \dfrac{1}{1-x(\tau)} \left[\left(3 - x(\tau)\right)\dfrac{\delta T}{T} \right.\right. \\
\left.\left. + \dfrac{\partial \delta T / \partial r}{\diff T / \diff r} - \dfrac{\delta\kappa}{\kappa} - \dfrac{\delta\rho}{\rho} - \dfrac{\partial\xi_r}{\partial r}\right]\right)~.
\label{eq:deltaL}
\end{multline}

In the following, we will adopt a solar-calibrated Hopf function, which is based on a numerical solution to the non-LTE radiative transfer equation derived by \citet{vernazza81} to match the observed solar spectrum. This Hopf function is furthermore consistent with 3D hydrodynamic atmospheric simulations provided by CO$^5$BOLD, which we later use to model the source term in Eq. \eqref{eq:wave_eq}. The Hopf function reads \citep[see][]{sonoi19}
\begin{equation}
q(\tau) = 1.036 - 0.3134\exp(-2.448\tau) - 0.2959\exp(-30.0\tau)~.
\label{eq:hopf}
\end{equation}

\begin{figure}
\centering
\includegraphics[width=\linewidth]{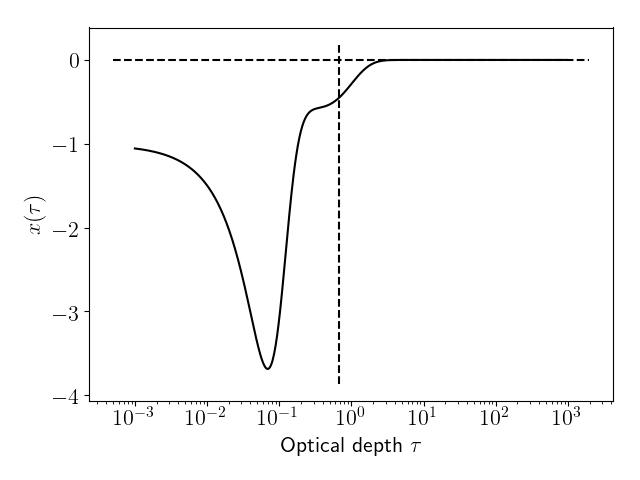}
\caption{Quantity $x(\tau)$ defined by Eq. \eqref{eq:xtau} as a function of optical depth $\tau$, when the Hopf function given by Eq. \eqref{eq:hopf} is adopted. The vertical dashed line corresponds to the radius of the photosphere ($\tau = 2/3$), where the intensity power spectrum is observed. The $\tau$-axis is oriented from left to right, so that the surface is on the left side of the figure, and the deeper layers on the right side.}
\label{fig:condition}
\end{figure}

In Fig. \eqref{fig:condition}, we plot the quantity $x(\tau)$ against $\tau$, where the Hopf function given by Eq. \eqref{eq:hopf} is adopted. As expected, $x(\tau)$ is much smaller than unity at high optical depth. This is in accordance with the condition that the Hopf function must reduce to a constant in the deeper layers of the star. At lower optical depth, however, $x(\tau)$ is no longer negligible. The height of observation of the continuum intensity power spectrum ($\tau \sim 2/3$, vertical dashed line in Fig. \ref{fig:condition}) lies halfway between these two extreme cases: we will therefore account for the quantity $x(\tau)$ in the following, even though it should not be of utter significance.

On a side note, comparing Eq. \eqref{eq:deltaL} to the perturbed radiative diffusion equation \citep[see Eq. 21.15 of][]{unno89}
\begin{equation}
\dfrac{\delta L}{L} = \dfrac{L_R}{L}\left(2\dfrac{\xi_r}{r} + 3\dfrac{\delta T}{T} - \dfrac{\delta \kappa}{\kappa} - \dfrac{\delta\rho}{\rho} + \dfrac{\partial \delta T / \partial r}{\diff T/\diff r} - \dfrac{\partial \xi_r}{\partial r}\right)~,
\label{eq:deltaL_diff}
\end{equation}
it can be seen that the intensity fluctuations of a grey atmosphere with an arbitrary Hopf function $q(\tau)$ reduce to their diffusive counterpart when the following condition is met:
\begin{equation}
x(\tau) \ll 1~.
\label{eq:condition}
\end{equation}
In particular, in the limit where the Hopf function is linear in $\tau$ -- another common approximation in stellar atmospheres, leading to the widely used Eddington-Barbier relations for instance --, $x(\tau)$ strictly vanishes, and the luminosity fluctuations of the grey atmosphere are strictly identical to the expression obtained in the diffusion approximation. Note, however, that this is only true of the fluctuating part of the luminosity. Even if $x(\tau)$ vanishes, the total luminosity of the grey atmosphere is not equal to the total luminosity obtained in the diffusion approximation.

\subsection{Relating the intensity fluctuations to the $\Psi_\omega$ variable}

Using the perturbed equation of state and the perturbed continuity equation, Eq. \eqref{eq:deltaL} can be rewritten in terms of fluid displacement and temperature fluctuations
\begin{multline}
\dfrac{\delta L}{L} = \dfrac{L_R}{L}\left(\left(2+\dfrac{2+3\kappa_\rho}{1-x}\right)\dfrac{\xi_r}{r} + \dfrac{4-x-\kappa_T}{1-x}\dfrac{\delta T}{T}\right. \\
\left. \vphantom{12pt} + \dfrac{r\kappa_\rho}{1-x}\dfrac{\diff (\xi_r/r)}{\diff r} - \dfrac{H_T}{1-x}\dfrac{\diff (\delta T/T)}{\diff r} \right)~,
\label{eq:L-T-xi}
\end{multline}
where $H_T \equiv |\diff \ln T / \diff r|^{-1}$ is the temperature scale height, $\kappa_\rho \equiv (\partial \ln\kappa / \partial\ln\rho)_T$, $\kappa_T \equiv (\partial \ln\kappa / \partial\ln T)_\rho$, and $x$ is the value of the quantity $x(\tau)$, defined by Eq. \eqref{eq:xtau}, at the considered optical depth.

One can then write the temperature fluctuations as a function of fluid displacement. While it is known that non-adiabatic effects are important at the top of the convection zone, they are considerably complicated to account for in the treatment of solar-like oscillations (in particular, including them in this formalism would require going beyond the second-order in Eq. \ref{eq:wave_eq}). As a first step, we therefore choose to neglect their impact on $p$-mode asymmetry, and we place ourselves in the quasi-adiabatic approximation. Further discussion regarding the validity of this assumption can be found in Sect. \ref{sec:non-adia}. In this approximation, the relative temperature fluctuations are proportional to the relative density fluctuations, which are related to the fluid displacement through the perturbed continuity equation. In turn, the fluid displacement is related to the velocity variable through $\widehat{\vv_\text{osc}}(r) = j\omega \xi_r(r)$, and the velocity variable to the variable $\Psi_\omega$ through Eq. \eqref{eq:v-psi}. This finally leads to
\begin{equation}
\dfrac{\delta L}{L} = \dfrac{L_R}{L}\left(\vphantom{\dfrac{L_R}{L}}A_L \Psi_\omega + B_L\dfrac{\diff \Psi_\omega}{\diff r} + C_L \dfrac{\diff^2 \Psi_\omega}{\diff r^2}\right)~,
\label{eq:L-psi}
\end{equation}
with
\begin{equation}
\begin{array}{l}
A_L = \left(2+\dfrac{2+3\kappa_\rho}{1-x}\right)A_\xi + \dfrac{4-x-\kappa_T}{1-x}A_T + \dfrac{r\kappa_\rho}{1-x}\dfrac{\diff A_\xi}{\diff r} - \dfrac{H_T}{1-x}\dfrac{\diff A_T}{\diff r} \\
B_L = \dfrac{4-x-\kappa_T}{1-x}B_T + \dfrac{r\kappa_\rho}{1-x} A_\xi - \dfrac{H_T}{1-x}\left(A_T + \dfrac{\diff B_T}{\diff r}\right) \\
C_L = -\dfrac{H_T}{1-x} B_T~,
\end{array}
\end{equation}
and
\begin{equation}
\begin{array}{l}
A_T = -\dfrac{\nabla_\text{ad}\Gamma_1}{r^2}\dfrac{\diff}{\diff r}\left(\dfrac{r}{c\sqrt{\rho}}\right) \\
B_T = -\dfrac{\nabla_\text{ad}\Gamma_1}{rc\sqrt{\rho}} \\
A_\xi = \dfrac{1}{r^2 c\sqrt{\rho}}~,
\end{array}
\end{equation}
where $\nabla_\text{ad}$ is the adiabatic gradient, and $\Gamma_1$ is the gas adiabatic index.

Eq. \eqref{eq:L-psi} shows that the intensity fluctuations cannot be described as being simply proportional to the wave equation variable $\Psi_\omega$, but also depend on its first and second derivatives. It can be seen, for instance from the last two terms in the right-hand side of Eq. \eqref{eq:L-T-xi}, that this is because the radiative flux is sensitive to the temperature gradient. In prior studies tackling the issue of asymmetry reversal, it is always either explicitly argued, or implicitly assumed, that the Green's function associated to intensity (or its pressure proxy) and velocity are proportional to one another \citep[see for instance][]{kumarB99}. Such a relation would, with the above notations, correspond to a case where only one of the coefficients $A_L$, $B_L$ or $C_L$ is non zero. In particular, synthetic spectrum models usually assume that the effective temperature fluctuations are equal to the actual temperature fluctuations, in which case only $B_L \neq 0$ (see Sect. \ref{sec:why} for a more detailed discussion).

\subsection{Intensity power spectral density}

The remainder of the procedure is identical to the one detailed in \citet{philidet20}. More precisely, to derive $\Psi_\omega$, we convolve the Green's function $G_\omega$ and the source term $S$ associated to the inhomogeneous wave equation (Eq. \ref{eq:wave_eq}). This is done by considering only the contribution of source localisation to the $p$-mode asymmetries: in particular, we do not account for the contribution of correlated noise, unlike what we did in our previous study. We also assume that the only source term in Eq. \eqref{eq:wave_eq} stems from fluctuations of the turbulent pressure. Finally, using the resulting $\Psi_\omega$ in Eq. \eqref{eq:L-psi} leads to
\begin{multline}
\left\langle\left|\widehat{\dfrac{\delta L}{L}}\right|^2\right\rangle = \left(\dfrac{L_R}{L}\right)^2 \left(\vphantom{\left\langle\left|\widehat{\dfrac{\delta L}{L}}\right|^2\right\rangle} A_L^2 I_\omega(X_\omega, X_\omega) + B_L^2 I_\omega(X'_\omega, X'_\omega) \right. \\
+ C_L^2 I_\omega(X''_\omega, X''_\omega) + 2A_L B_L I_\omega(X_\omega, X'_\omega) + 2A_L C_L I_\omega(X_\omega, X''_\omega) \\
\left. \vphantom{\left\langle\left|\widehat{\dfrac{\delta L}{L}}\right|^2\right\rangle} + 2B_L C_L I_\omega(X'_\omega, X''_\omega)\right)~,
\label{eq:1}
\end{multline}
where $\langle\rangle$ denotes ensemble average, and the function $I_\omega$ is defined as
\begin{multline}
I_\omega(f_1,f_2) = \displaystyle\iint \diff^3\bm{r_{s1}}\diff^3\bm{r_{s2}} \mathrm{Re}\left(\partial_i f_1(\bm{r_{s1}})\partial_j f_2^\star(\bm{r_{s2}})\right) \\
\times\left\langle\left.(\rho_0\widehat{u_r u_i})\right|_{\bm{r_{s1}}} \left.(\rho_0\widehat{u_r u_j}^\star)\right|_{\bm{r_{s2}}}\right\rangle~,
\label{eq:2a}
\end{multline}
where the Einstein convention on index summation is used for the indices $i$ and $j$, $^\star$ denotes complex conjugation, $\mathrm{Re}$ denotes the real part of a complex quantity, and $X_\omega$, $X'_\omega$ and $X''_\omega$ are defined in terms of the Green's function as
\begin{equation}
\begin{array}{l}
X_\omega(\bm{r_s}) = \dfrac{r}{c\sqrt{\rho}}G_\omega(\bm{r_o};\bm{r_s})~, \\
X'_\omega(\bm{r_s}) = \dfrac{r}{c\sqrt{\rho}}\left.\dfrac{\partial G_\omega}{\partial r_o}\right|_{\bm{r_o};\bm{r_s}}~, \\
X''_\omega(\bm{r_s}) = \dfrac{r}{c\sqrt{\rho}}\left.\dfrac{\partial^2 G_\omega}{\partial r_o^2}\right|_{\bm{r_o};\bm{r_s}}~,
\end{array}
\label{eq:2b}
\end{equation}
where the variables $r_o$ and $r_s$ in the Green's function refer to the radius of observation and the radius of the source respectively. We note that the coefficients $A_L$, $B_L$ and $C_L$ in Eq. \eqref{eq:1} are only evaluated at the height of observation of the modes. As such, and as we pointed out above, we only need to model the oscillation-induced radiative flux variations at the photosphere, whereas the effect of the spatial extent of the source is entirely contained within the function $I_\omega$. In particular, this is the reason why the value of the quantity $x(\tau)$ is only important at the height of observation (illustrated by the vertical dashed line in Fig. \ref{fig:condition}).

It is worthwhile to mention that Eq. \eqref{eq:1} does not involve any further modelling than when the velocity power spectrum is calculated. Indeed, except for additional equilibrium thermodynamic quantities, both the Green's function and the fourth-order correlation term in Eq. \eqref{eq:2a} were already modelled in \citet{philidet20}. Otherwise stated, this means that modelling the contribution of source localisation to asymmetries in intensity does not require additional physical constraints, nor does it require new input parameters to be introduced, provided the non-adiabatic terms, not accounted for in this model, indeed remain small in the stellar superficial layers.

\begin{figure*}
\centering
\includegraphics[scale=0.8]{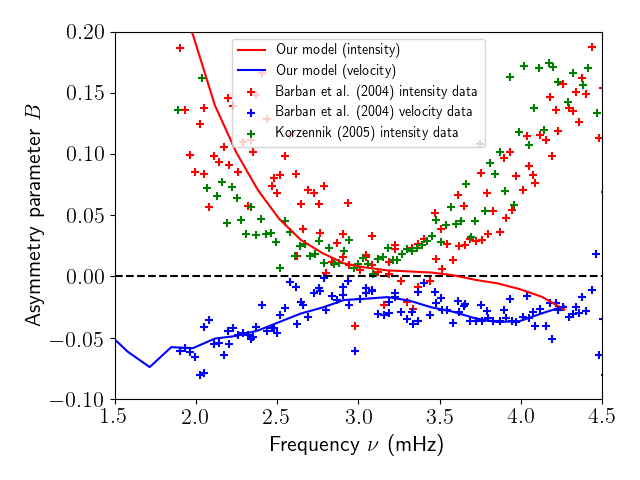}
\caption{Asymmetry profile $B(\nu)$ obtained by the ``numerical spectrum'' model with $\lambda = 0.5$, in the intensity power spectrum (\textit{solid red line}) and in the velocity power spectrum (\textit{solid blue line}). Both are compared to the asymmetry profile $B(\nu)$ inferred from observations by \citet{barban04} in the intensity (\textit{red crosses}) and the velocity power spectra (\textit{blue crosses}) respectively. The green crosses correspond to the asymmetry measured by \citet{korzennik05} in the intensity power spectrum.}
\label{fig:obs}
\end{figure*}

\section{Results for intensity asymmetries and comparison with observations\label{sec:obs}}

\subsection{Numerical computation of intensity asymmetries}

The system comprised of \Cref{eq:1,eq:2a,eq:2b} constitutes a closed, semi-analytical form of the intensity power spectrum. The equilibrium quantities involved, as well as the turbulent quantities needed as inputs for the analytical model of turbulence, are all extracted from a 3D hydrodynamic simulation of the solar atmosphere computed using the CO$^5$BOLD code, which, after horizontal and time average, we patched on top of a 1D solar model provided by the evolutionary code CESTAM \citep[for more detail, see][]{manchon18}. Once the intensity power spectrum is calculated, we fit the obtained synthetic line profiles using the following formula \citep{nigamK98}
\begin{equation}
P(\omega) = H_0\dfrac{(1+Bx)^2+B^2}{1+x^2}~,
\label{eq:defb}
\end{equation}
where $x = 2(\omega-\omega_0)/\Gamma_{\omega_0}$ is the reduced pulsation frequency, $\omega_0$ the angular eigenfrequency and $\Gamma_{\omega_0}$ the linewidth of the mode. The parameter $B$ corresponds to the asymmetry parameter. The terms \textit{positive} and \textit{negative asymmetry} refer to the sign of $B$, and in the special case $B=0$ we recover a Lorentzian line profile.

In order to compare our results to observational data, we used the numerical spectrum model developed in \citet{philidet20}. It contains only one parameter in the form of $\lambda$ (see their Sect. 2.3.2 for a definition), which illustrates the relative uncertainty pertaining to the temporal spectrum associated to the turbulent cascade in the solar superficial layers. Physical arguments allow to constrain the value of this parameter \citep[see for instance][]{samadiG01}; to be consistent with these constraints, they retained the value $\lambda = 0.5$. This led to a satisfactory quantitative agreement between their model and observations in the velocity spectrum. Here, we retain the same value for $\lambda$, and compare the resulting intensity asymmetries to observations: there is therefore no parameter adjustment in what follows.

\subsection{Observational datasets}

To compare the asymmetries predicted by the model with observations, we use the same observational data as for the velocity power spectrum in \citet{philidet20}. These data points were extracted from the spectrum analysis conducted by \citet{barban04} on observations made by the \textit{GONG} network. We used these observations because they are particularly fitting for the study of asymmetry reversal: indeed, the \textit{GONG} network provides with velocity and intensity measurements simultaneously. We recall that the spectral analysis of \citet{barban04} considers non-radial modes ($15 < l < 50$), whereas our model was developed for radial modes. However, the dependence of asymmetry on angular degree is very weak, and the asymmetry observations collapse to a slowly varying function of frequency, so long as $l \lesssim 100$ \citep[see e.g][]{vorontsovJ13}. Therefore, the frequency-dependence of asymmetry for the modes observed by \textit{GONG} is expected to be almost identical to that of radial modes.

To obtain the asymmetries in the intensity spectrum, \citet{barban04} used approximately one year of \textit{GONG} data, allowing to resolve about $600$ modes. Following \citet{severino01}, they fitted simultaneously the velocity spectrum, intensity spectrum, and I-V cross spectrum (both coherence and phase difference), which is known to yield more reliable results. As such, the observed asymmetries we used in velocity and intensity are not obtained independently, but through the same fitting procedure. The model used for the spectral analysis, however, considered that the asymmetry was entirely due to a coherent non-resonant background component, so that their results were presented in terms of noise level and phase differences. This is a different mathematical description for the same line profile shape as given by Eq. \eqref{eq:defb}. In order to extract the asymmetry parameter $B$ as defined by Eq. \eqref{eq:defb}, we reconstructed the line profiles using the best-fit values obtained by \citet{barban04}, and then fitted these reconstructed line profiles using Eq. \eqref{eq:defb} instead.

We also consider intensity asymmetries inferred from \textit{HMI} observations, and extracted from the spectral analysis procedure described in \citet{korzennik05}. Indeed, the asymmetry profile $B(\nu)$ resulting from this spectral analysis, while in agreement with the values inferred by \citet{barban04}, shows less dispersion. As such, it allows for a more robust comparison of our results with observations. The dataset used contains 4 periods of 72 days each. In \citet{korzennik05}, the asymmetries are already directly given in terms of a parameter $\alpha$ which is easily related to the parameter $B$, through $B = \alpha / 2$. We only considered angular degrees between $l=0$ and $l=20$, and we averaged the values of $B$ over bins of width $30~\mu$Hz.

\subsection{Modelled asymmetries compared with observations}

Our model allows us to predict the mode asymmetries in both the velocity and intensity spectra simultaneously. However, since we have already focused on the results pertaining to the Sun's velocity power spectrum in \citet{philidet20}, we will focus the subsequent analysis on the Sun's intensity power spectrum.

Fig. \ref{fig:obs} showcases the comparison between the asymmetries predicted by our model and those inferred from observations, both in velocity and intensity. It is clear that our model predicts a reversal of the asymmetries between these two observables, except for the higher order modes ($\nu \gtrsim 3.6$ mHz). It is also clear that a quantitative agreement is found with observations for modes with $\nu \lesssim \nu_\text{max}$, where $\nu_\text{max} \sim 3$ mHz is the frequency of maximum height in the $p$-mode power spectrum. This shows that the localisation of the source, which is the primary cause of asymmetry in the velocity spectrum, is sufficient to explain a large portion of the asymmetry reversal.

However, our model fails to account for the intensity asymmetry of the higher-order modes, with $\nu \gtrsim \nu_\text{max}$. Indeed, while it consistently predicts slightly negative asymmetries for these modes, observations tend to show that they actually feature strong positive asymmetry. This shows that other mechanisms must be invoked to fully explain the asymmetry reversal throughout the entire $p$-mode spectrum. For instance, unlike the velocity power spectrum, our intensity power spectrum model does not contain the contribution of the correlated background. It is rather consensual that this contribution is negligible in velocity data. In turn, the results presented in Fig. \ref{fig:obs} indicate that it is also negligible in intensity data for low frequency modes, but that it may no longer be the case for high frequency modes. Furthermore, the non-adiabaticity of the oscillations is not taken into account in our study: we discuss this approximation in Sect. \ref{sec:non-adia}. The opacity effect -- as it was coined by \citet{severino08} -- could also play a role. Finally, the fluctuations of the convective flux may add a non negligible contribution to the total intensity fluctuations, whereas we only considered the fluctuations of the radiative flux in Eq. \eqref{eq:deltaL_diff}. The relative importance of both kind of fluxes depends on the height at which the modes are observed. In particular, the higher the radius of observation, the more prominent the radiative flux.

\section{Discussion\label{sec:discussion}}

\subsection{Understanding the asymmetry reversal puzzle\label{sec:why}}

As already mentioned in Sect. \ref{sec:model}, the reason why source localisation leads to different asymmetries in the intensity and velocity power spectra is the fact that the relation which links the intensity fluctuations to the velocity fluctuations (or, equivalently, to the variable $\Psi_\omega$) is not a simple linear relation, but involves its first and second derivatives as well. As we also mentioned above, this is because the radiative flux depends primarily on the temperature gradient, and not only on the absolute value of the temperature. As such, there can be partial cancellation between the three terms on the right-hand side of Eq. \eqref{eq:L-psi}, leading to line profiles that are not simply proportional to each other.

In contrast, using simpler approximations to model the intensity fluctuations leads to much simpler relations between $\delta L/L$ and $\vv_\text{osc}$, which, precisely because of their simplicity, are unable to predict any asymmetry reversal. In the following, we illustrate this fact by considering that $\delta T_\text{eff} / T_\text{eff} = \delta T / T$ at the photosphere. In that simple case, the intensity fluctuations can be written
\begin{equation}
\dfrac{\delta L}{L} = 4\dfrac{\delta T}{T} + 2\dfrac{\xi_r}{r}~.
\label{eq:bb}
\end{equation}
It is readily shown that the second term in the right-hand side of Eq. \eqref{eq:bb} is negligible for solar $p$-modes close to the surface. In turn, the temperature fluctuations are, in the quasi-adiabatic approximation, proportional to the pressure fluctuations. In this simple model, pressure fluctuations are therefore a suitable proxy for intensity fluctuations. Then, using the perturbed continuity equation, the intensity fluctuations are related to the variable $\Psi_\omega$ through a relation similar to Eq. \eqref{eq:L-psi}, but with
\begin{equation}
\begin{array}{l}
B_L = -\dfrac{4\nabla_\text{ad}\Gamma_1}{rc\sqrt{\rho}} ~, \\
\vspace{0.5pt} \\
A_L = C_L = 0~.
\label{eq:L-psi-bb}
\end{array}
\end{equation}

\begin{figure}
\centering
\includegraphics[width=\linewidth]{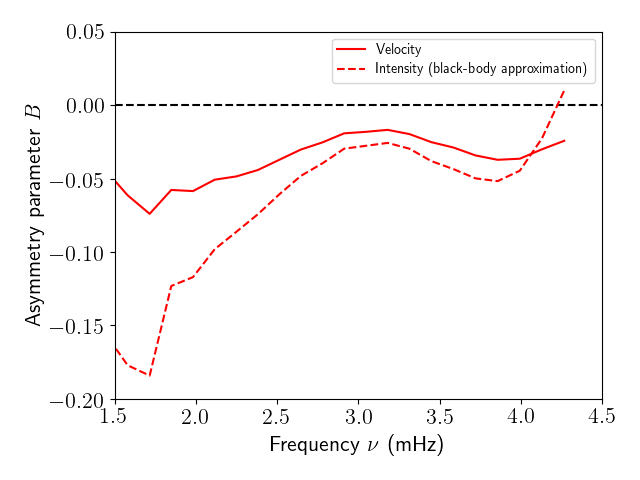}
\caption{Predicted asymmetry profiles $B(\nu)$ when the relation $\delta T_\text{eff} = \delta T$ is assumed. The solid red line shows the velocity asymmetries (since they are unaffected by the intensity fluctuations modelling, it is identical to the solid blue line in Fig. \ref{fig:obs}). The dashed red line corresponds to the intensity asymmetries obtained by using Eq. \eqref{eq:bb} in place of Eq. \eqref{eq:L-psi}.}
\label{fig:bb_vs_rad}
\end{figure}

We compare in Fig. \ref{fig:bb_vs_rad} the velocity and intensity asymmetry profiles $B(\nu)$ obtained when the relation $\delta T_\text{eff} / T_\text{eff} = \delta T / T$ is adopted at the photosphere. As can readily be seen, the asymmetries are not reversed. This is because in this simplified model, only one term is retained in the relation between the two types of fluctuations.

The relation between the intensity and velocity Green's functions is therefore a crucial key to tackling the issue of asymmetry reversal. In many studies on the subject, various approximations were made to model this relation. For instance, when dealing with asymmetry reversal, \citet{duvall93} split the wave variables into dynamical variables, such as velocity, and thermal variables, such as temperature or brightness. They then implicitly consider that each set of variables is characterised by one Green's function, and in particular that temperature and intensity fluctuations are proportional to one another, which is analogue to Eq. \eqref{eq:bb}. Adopting Eq. \eqref{eq:bb}, as shown by \citet{dupret02}, does not substantially alter predictions regarding intensity fluctuations, provided temperature fluctuations are described in a fully non-adiabatic framework. However, the use of a simple Newton's cooling law to model non-adiabaticity renders it invalid. \citet{rastB98,nigam98,kumarB99} made similar assumptions regarding the intensity fluctuations, all of which are equivalent to considering the approximation $\delta T_\text{eff} / T_\text{eff} = \delta T / T$, without the necessary non-adiabatic framework.

On the other hand, \citet{roxburghV97} used the same Green's function for velocity and intensity, but considered different source terms. More specifically, they added a frequency-independent source term for the velocity fluctuations, which they did not consider for the intensity fluctuations. In other words, they considered that the only difference between the velocity and intensity fluctuations is the addition of a coherent, non-resonant background component in the velocity signal. Likewise, in \citet{chaplinA99}, the authors considered proportional Green's functions for velocity and intensity. All these assumptions are equivalent to considering that the coefficients $B_L$ and $C_L$ in Eq. \eqref{eq:L-psi} are zero, thus only retaining the coefficient $A_L$.

Prior studies on asymmetry reversal thus have this in common, that they use simplifying approximations to model the intensity fluctuations, so that their equivalent of Eq. \eqref{eq:L-psi} only contains one non-vanishing coefficient (either $A_L$ or $B_L$, depending on the authors). This is the reason why, in their models, source localisation alone could not explain the different sense of asymmetry observed in velocity and intensity.

\subsection{The quasi-adiabatic approximation\label{sec:non-adia}}

As we mentioned above, our model describes the oscillations in the quasi-adiabatic framework, in the sense that, in establishing the wave equation given by Eq. \eqref{eq:wave_eq}, we neglect all heat transfers, both in the homogeneous part -- the full, non-adiabatic wave equation is of fourth order, whereas ours is only second-order -- and in the source term -- we only considered mechanical work exerted by fluctuations of the turbulent pressure, and left out the effect of turbulent entropy fluctuations. As pointed out by \citet{gabriel98}, increasing the order of the wave equation is likely to cause asymmetries in different observables to drift further apart. As such, in order to predict the shape of the mode line profiles as realistically as possible, one should go beyond the quasi-adiabatic approximation, and perform fully non-adiabatic computations. However, as a first step, we adopted this approximation in this study.

The quasi-adiabatic approximation is notoriously questionable in the superficial layers of solar-like oscillators, and more specifically in the super-adiabatic region \citep{samadireview,houdekreview}, because the thermal timescale -- over which heat transfers typically occur -- in this region is neither much smaller, nor much greater than the period of the oscillations, but rather coincide with it. As such, modal entropy fluctuations are non-zero, and should, in a fully non-adiabatic framework, be included in the equation of state -- along with their corresponding evolution equation. The difference shows, for example, in the phase difference between velocity and intensity modal fluctuations: while the quasi-adiabatic approximation yields a phase difference of $\pi / 2$ between the two, the observed phase difference varies significantly with frequency \citep{barban04}.

Therefore, in adopting the quasi-adiabatic approximation, we do not overlook its shortcomings, but rather consider this a first step to a novel approach, which is still likely to shed light on the issue of asymmetry reversal. The impact of non-adiabaticity is perhaps best illustrated with mode amplitudes. Indeed, in discarding modal entropy fluctuations from the oscillation model, predicted intensity amplitudes are severely overestimated compared to their observed counterpart.

\begin{figure}
\centering
\includegraphics[width=\linewidth]{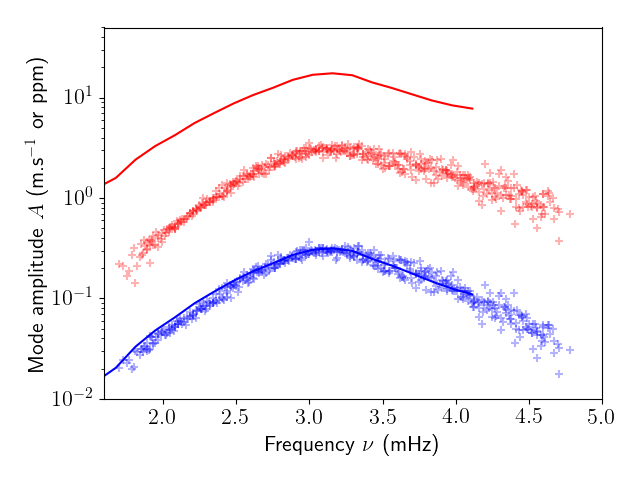}
\caption{$p$-mode amplitudes $A$, as defined by Eq. \eqref{eq:amp}, as a function of mode frequency $\nu$. The legend is identical to that of Fig. \ref{fig:obs}: crosses correspond to amplitudes inferred from observations, using the spectral analysis of \citet{barban04} (each cross represents an individual doublet $(n,l)$), whereas solid lines correspond to the prediction of our model; blue element pertain to velocity, red elements to intensity. The amplitudes are given in m.s$^{-1}$ for velocity, and in ppm for intensity.}
\label{fig:amps}
\end{figure}

This is illustrated in Fig. \ref{fig:amps}, where we show the amplitudes predicted by our model as well as the observed solar mode amplitudes, both for velocity and intensity. Predicted mode amplitudes are accessed from the same fitting procedure described in Sect. \ref{sec:obs}; the amplitude squared of a mode indeed corresponds to the area under the curve of its spectral power density, and is therefore given by

\begin{equation}
A^2 = \pi\Gamma_\omega H_0~,
\label{eq:amp}
\end{equation}
where $A$ is the mode amplitude, $\Gamma_\omega$ its linewidth and $H_0$ is defined by Eq. \eqref{eq:defb}. Note that mode amplitudes defined thusly are intrinsic amplitudes, and do not account for visibility factors. As for the observed amplitudes, we use the results of the same spectral analysis conducted by \citet{barban04} that we used to compare asymmetries in Sect. \ref{sec:obs}.

It can be seen from Fig. \ref{fig:amps} that while our model reproduces the amplitudes in velocity to a satisfactory extent, the intensity amplitudes are overestimated by an order of magnitude (approximately a factor $5$). In particular, it appears that the overestimation factor only slightly depends on frequency. This overestimation is a known shortcoming of the quasi-adiabatic approximation, whose alleviation, as we mentioned above, would require going beyond a second-order wave equation. We note, however, that simply because non-adiabatic effects have an important impact on mode amplitude does not mean that the impact on asymmetries is as critical. Indeed, asymmetries are only sensitive to differential, frequency-dependent alterations of the line profiles, and remain unchanged under an overall reduction factor. An alternative approach which may be pursued would be to obtain an equivalent of Eq. \eqref{eq:deltaTeff} by replacing $\delta\tau / \tau$ in Eq. \eqref{eq:deltaTsurT} by its integral expression obtained through Eq. \eqref{eq:deltatausurtau}. Since this method does not rely on $\tau$-derivatives, it is likely to be less affected by departure from adiabaticity; however, the integral form of $\delta\tau / \tau$ renders the resulting equivalent of Eq. \eqref{eq:1} impractical to use.

\section{Conclusion}

In this paper, we extended the semi-analytical synthetic power spectrum model developed in \citet{philidet20} to predict the radial $p$-mode asymmetries in intensity observations. We treated the $p$-mode-induced fluctuations of the radiative flux by perturbing a grey atmosphere model with a solar-calibrated Hopf function. We showed that the dependence of the radiative flux on the temperature gradient, and not simply on the absolute value of the temperature, has a profound impact on the shape of the line profiles in the intensity spectrum. We find quantitative agreement between our predicted asymmetries and the corresponding observations for low-frequency modes, both in velocity and intensity simultaneously. We conclude that no secondary physical mechanism is necessary to explain the reversal between asymmetries in velocity and intensity, at least for $\nu \lesssim \nu_\text{max}$.

Our model is unable, however, to explain asymmetry reversal for higher-order modes. Other, secondary mechanisms can be invoked there: the non-adiabatic effects invoked by \citet{duvall93}, the correlated background effect invoked by \citet{nigam98}, or the impact of the convective flux fluctuations, are all viable candidates. The fact that these discrepancies are restricted to high-frequency modes in intensity suggests that non-adiabaticity is at least partly responsible. As a concluding remark, we point out that radiative transfers, even if they are not sufficient to reverse asymmetries in the high-frequency end of the $p$-mode spectrum, still have a significant impact on the line profiles: as a result, in investigating the impact of any other physical mechanism on asymmetry reversal, this intrinsic difference between velocity and intensity line profiles must still be accounted for.

The main limitation of the model presented in this study is the quasi-adiabatic approximation, which is not valid in the superficial layers of the star. Although it sheds some light on the asymmetry reversal problem, any further improvement of this model therefore will necessitate going beyond this approximation and considering a non-adiabatic wave equation instead. We postpone this refinement to a later study.

\begin{acknowledgements}
The authors wish to thank the anonymous referee for his/her useful comments, which helped considerably improve this manuscript. J.P, K.B and R.S acknowledge financial support from the `Programme National de Physique Stellaire' (PNPS) of CNRS/INSU and from the `Axe Fédérateur Etoiles' of Paris Observatory. H.G.L. acknowledges financial support by the Sonderforschungsbereich SFB\,881 ``The Milky Way System'' (subprojects A4) of the German Research Foundation (DFG).
\end{acknowledgements}

\bibliographystyle{aa}
\bibliography{letter_biblio}

\end{document}